# Carving Nature/Conceptual Models at Joints Using Thinging Machines

**Sabah Al-Fedaghi***
*Computer Engineering Department*
*Kuwait University*
*Kuwait*

***Abstract - To handle the complexity of our world, the carving metaphor has been used to build a conceptual system of reality. In such an endeavor, we can choose various joints to carve at; that is, we can conceptualize various aspects of reality. Conceptual modeling concerns carving (e.g., categorization) and specifying a conceptual picture of a subject domain. This paper concerns with applying the notion of carving to conceptual models. Specifically, it concerns modeling based on the so-called thinging machine (TM). The central problem is how to carve events when building a TM model. In TMs, an event is defined as a thimac (thing/machine) with a time feature that infuses dynamism into the static thimac, called a region. A region is a diagrammatic description based on five generic actions: create, process, release, transfer, and receive. The paper contains new material about TM modeling and generalization and focuses on the carving problem to include structural carving and dynamic events. The study's results provide a foundation for establishing a new type of reality carving based on the TM model diagrams.***

## I. INTRODUCTION

Plato employed the carving metaphor as an analogy for a world comes as a pre-partitioned world. Such a notion has been used in building a conceptual scheme that involves organizing constructs through which reality is perceived and organized. In such an endeavor, we can choose various joints to carve at; that is, we can conceptualize various aspects of reality.

Conceptual modeling is a field that concerns such a topic and how to specify and develop a conceptual description of a subject domain. Conceptual models are used to support the design of databases, software, business processes, enterprises, etc. For example, suppose that we want to navigate a given city by using the subway system. By using notions such as subway line, subway station, line direction, ordering of stations, and intersection between lines, we can produce a conceptual representation of those particular subway lines [1].

According to [2],

> A conceptual system can succeed or fail to fit well, that is, to "carve nature at the joints." In addition, conceptual systems can vary in their "fineness of grain," that is, they can carve nature into big chunks or small artful slices. […] A conceptual system is accurate if it always finds joints in nature and inaccurate if it misses the joints and hits a bone or nothing at all.

In this context, the "joints" are between

(a) Things and their relationships, and
(b) Happenings and their relationships [3].

Note that point (b) is an important aspect in this study.

To build such a conceptual description, Plato employed a carving metaphor as an analogy for the reality of *Forms*: "Like an animal, the world comes to us predivided." Ideally, our best theories will be those which "carve nature at its joints" [4]. How is this carving actually achieved? *Categories* mark divisions between things that determine the basic categorical structure of the world. Speaking metaphorically, the fundamental *categorical* structure of the world carves the world at its fundamental joints [5].

### A. Categorization is a central issue

According to [4], the vocabulary of science aids in conceptualization (conceptual model of a *real* system) and communication. It also supports the process of "grouping particular things on the basis of shared properties, regularities, dispositions, natural laws, and so forth enables understanding and control" [4]. Therefore, a *categorization* of what is "found in or made by nature" emerges. This science of classification is a fundamental and dynamic science, and it concerns the basis of natural order [6]. Here, *naturalness* is often found in the core lexicon of natural languages, meaning that many languages have words that (roughly) correspond to *natural* concepts (e.g., animal species) [7].

Categories regarding the traditional view are characterized solely by the properties their members share [2]. Humans and many animals categorize things by branding various objects as members of instances categories [2]. Aristotle divided "the realm of being into at least two fundamental categories: the fundamental category of particulars, or the present-in, and the fundamental category of universals, or the said-of. This gives the world a certain sort of structure, built from things with two different natures" [5].

----------------------

* Retired June 2021, seconded fall semester 2021/2022

*B. Aims*

We apply this notion of carving in the context of recent research called TM modeling, a promising new approach to specifying conceptual systems. TM modeling is based on diagrammatic representation and meant to be applied in software engineering, just like the entity–relationship model and UML. TM modeling is unique because it does not suffer from mixing static constructs with dynamic events.

A central problem in this study is how to carve *events*. According to [3], representing knowledge about *things* in a subject domain and describing the *actions* within that domain are two fundamentally different tasks that need to be addressed separately but with an understanding of their interdependence.

In a TM, an event is defined in terms of a thimac (thing/machine) with a time aspect that infuses dynamism into the static description (called a region) of the thimac. A region is a diagrammatic specification based on five generic actions: create, process, release, transfer, and receive.

To illustrate the notion of carving *TM events*, consider the TM diagram in Fig. 1 (top), which models *A student going to school*. Classical categorization concerns carving such entities as *student* and *school*. In a TM, we are concerned, additionally, with carving "processes" or events such as *going-to*, *leaving*, and *arriving-at* (Fig. 1 (bottom)).

In this paper, we will not present a solution for the problem of carving static TM model into events; rather, we generalize, focus, and relate the carving problem to include processes and events as well as their combinations. The study's results provide a foundation for developing a new type of carving reality based on graphs based on the TM model diagrams.

*C. Sections*

For the sake of a self-contained paper, the next section includes a brief review of the TM model's theoretical foundation. Section 3 provides two illustrative examples of TM modeling. Section 4 includes reflections that clarify the ontological foundation of TM modeling. Section 5 includes a discussion about the notion of carving nature. Section 6 presents a case study that discusses the curving phenomena utilizing modeling patients in a hospital and their associated diagnoses.

## II. TM Modeling

This section includes a summary of the TM model discussed in previous papers (especially recent publications, e.g., [8]). In contrast to a world comprising substantial things and relationships, as philosophers have overwhelmingly supposed, TM modeling aims at developing a conceptual apparatus for reality based on thimacs. Thimacs are basically *form*s and *realization*s of processes constructed to include five actions. Their (static) form is a diagrammatic "space" that specifies boundaries and the structure of reality in terms of actions.

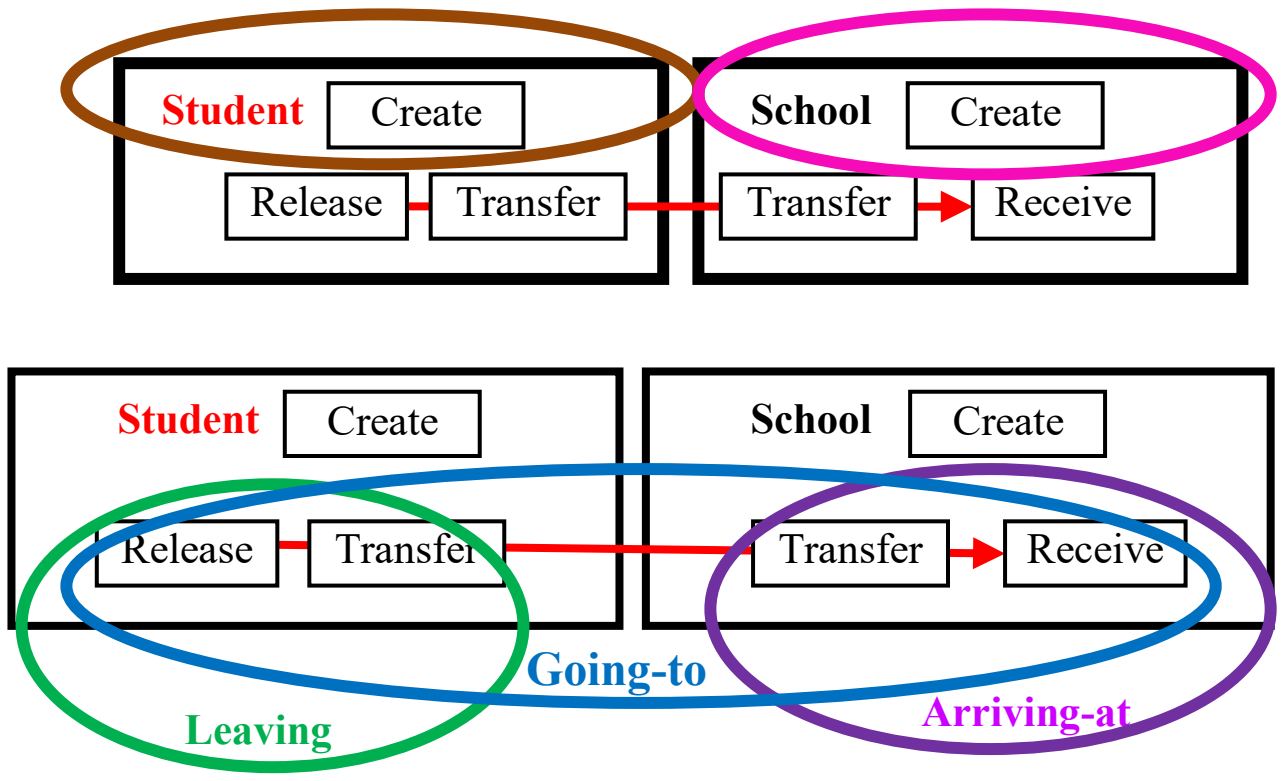

Fig. 1 Carving the events going-to, leaving from, and arriving at.

*A. Basic Description*

The TM model provides us with an ontological representation of reality. It conceptualizes the entities and processes (events) as two modes of one notion, which we call a *thimac* (*thi*ng/*mac*hine). Thimacs are defined in terms of five actions (Fig. 2). Things, numbers, sets, concepts, and propositions are thimacs.

Thimacs are "discriminative constructs" of the world. The world is divided into thimacs, and various thimacs overlap or combine to form the texture of the whole as a grand thimac. A generic thimac is a gathering up of elements into a unity or synthesis of actions: create, process, release, transfer, and receive. The thimac's constituents are formed from the makeup of these actions. An action is a unit of actionality. A TM diagram or sub-diagram is called a *region* at the static-modeling level. Fig. 3 shows an ontological picture that outlines the two levels of the TM scheme.

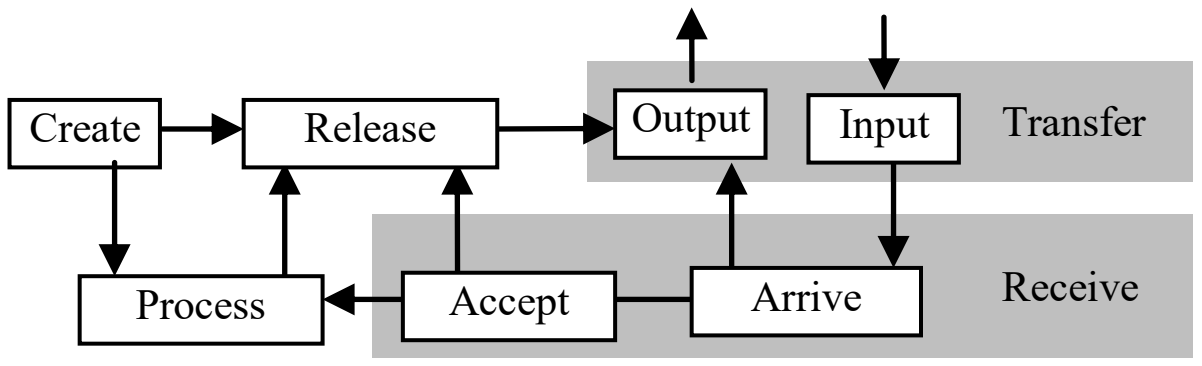

Fig. 2. Thimac.

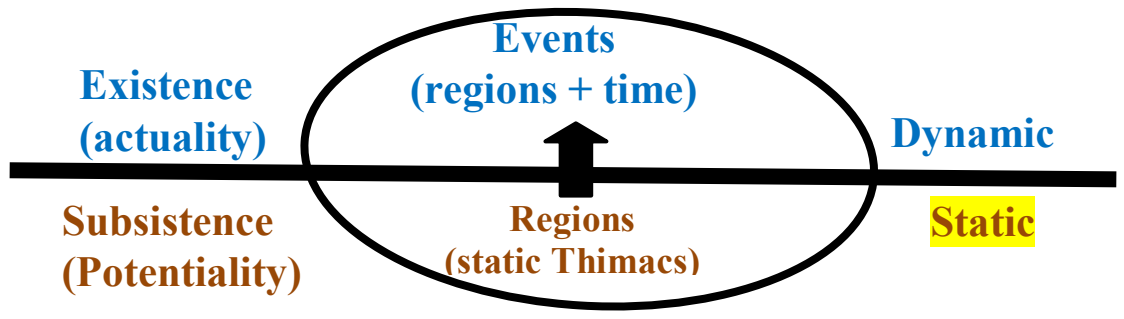

Fig. 3 Two levels of TM modeling.

The synthesis of actions is applied to *events* (at the dynamic level) (see Fig. 3). In a TM, there is no ontological distinction between concrete and conceptual things (e.g., mathematical concepts), and all are things. A thing is what can be created, processed, released, transferred and received. Additionally, the thimac is a machine that creates, processes, releases, transfers, and receives things. All so-called entities, properties, and relationships are thimacs or subthimacs, resulting in a "mesh," the interconnectedness of all things [9] of (a partial) world with one big complex thimac that contains all within it.

### B. *Actions*

At the most basic explanatory level, we conceive the thimac as constituted by two kinds of capacity, which we label *thing* and *machine*. A thing constitutes the thimac's capacity to be affected, whether by itself (i.e., one "part" or capacity of the thimac affecting another) or some other thimac (machine). In contrast, the power of machinery is an ability to initiate its actions from itself on other thimacs (things).

The thimac is a thing with a "structure" formed by the flow of other thimacs through it. It is a machine with five actions operates on things. A thimac's actions, shown in Fig. 2, are described as follows:

1) ***Arrive:*** A thing arrives at a thimac.
2) ***Accept:*** A thing enters a thimac. For simplification, the arriving things are assumed to be *accepted* (see Fig. 2); therefore, *arrive* and *accept* actions are combined into the ***receive*** action.
3) ***Release:*** A thing is ready for transfer outside the thimac.
4) ***Process:*** A thing is changed, handled, and examined, but no new thimac is generated.
5) ***Transfer:*** A thing crosses a thimac's boundary as input or output.
6) ***Create:*** *Creation* here refers to producing a thing that is *new to the world* (of the model); therefore, a new thimac is registered as an ontological unit. It indicates the birth or coming-into-thimac (i.e., must make it out of other thimacs). At the static level, *create* is a logical possibility of realization at the dynamic level.

Additionally, the TM diagrammatic model includes *storage* (represented as a **cylinder** in the TM diagram) and *triggering* (denoted by a **dashed arrow**). Triggering transforms from one series of movements to another (e.g., electricity triggers heat generation).

## III. ILLUSTRATIVE EXAMPLE

Before the topic of carving domains in TM modeling is discussed, this section includes two illustrative examples of TM modeling.

### A. *The circulatory system*

According to [10], the main human transport system is the circulatory system, a system of tubes (blood vessels) with a pump (the heart) and valves to ensure the one-way flow of blood (see Fig. 4). Its functions are

- To transport **nutrients** and **oxygen** to the cells.
- To remove **waste** and **carbon dioxide** from the cells.
- To provide for efficient **gas exchange**.

The **right** side of the heart collects **deoxygenated** blood from the body and pumps it to the **lungs**. The **left** side collects **oxygenated** blood from the lungs and pumps it to the **body**.

*Static TM Model*

Fig. 5 shows the TM static model of this circulatory system. In the figure,

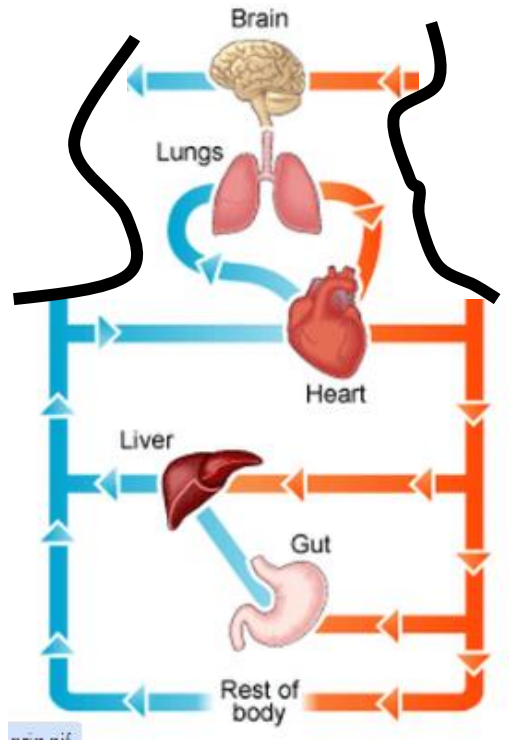

Fig. 4 Human blood circulatory system (from [10] – The picture is cut to achieve partial copying while giving the reader a general idea of the involved visual representation.

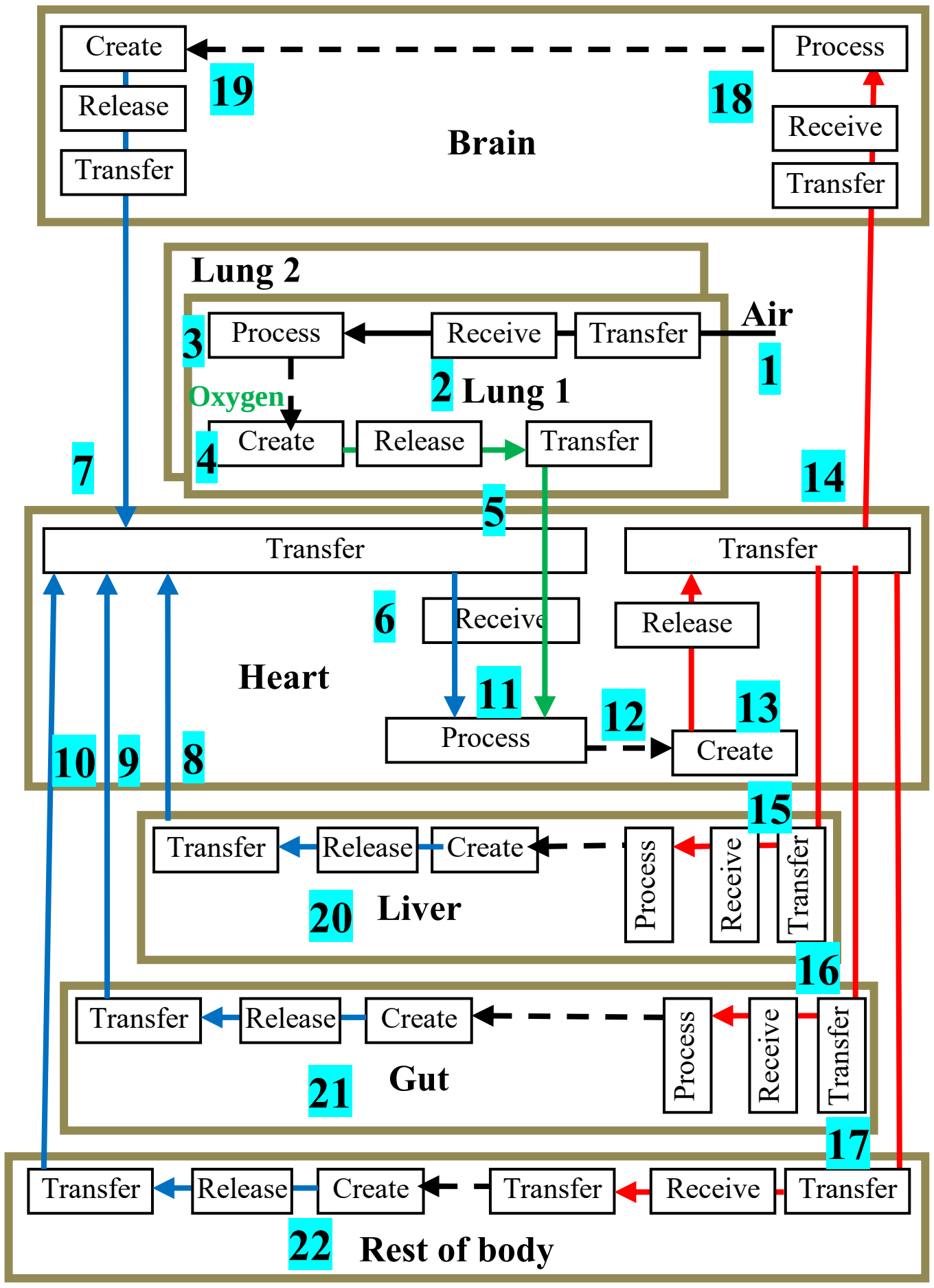

Fig. 5 Static TM model.

- Air flows into the lungs (notice that there are two lungs) to be processed (3) and triggers the appearance (manifestation) of oxygen (4).
- The Oxygen flows to the heart (5). Additionally, the heat receives (6) the deoxygenated blood from the brain (7), liver (8), gut (9), and the rest of the body (10). In the heart, the air and the incoming deoxygenated blood is processed (11) to trigger (12) the creation of processed oxygenated blood (13).
- The oxygenated blood flows to the brain (14), liver (15), gut (16), and the rest of the body (17).
- In the brain, the incoming oxygenated blood is processed (18) to create deoxygenated blood (19), which flows to the heart (7).
- The flows in the liver (20), gut (21), and the rest of the body (22) follow similar procedures to the one in the brain.
-

*Dynamic Model*

The dynamic model is superimposed on the static model using the notion of *event*. An event is a thimac that includes time. For example, Fig. 6 shows the event *Air enters lungs*. Note that the *region* in Fig. 6 is a sub-diagram of the static model. For simplicity's sake, regions will represent events in the dynamic model. Additionally, an individual event is "limited" by its region in contrast to such things as space, which is usually is said to be limited by that which it surrounds. The notion of limit here is important because only an event that has a limit can appear as an individual event, standing out from the multitudes of events that compose the background-as-stage of that appearance [11].

Accordingly, the following events are selected as shown in Fig. 7:

$E_1$: Air enters the lungs.
$E_2$: The air is processed to produce oxygen.
$E_3$: Oxygen flows to the heart.
$E_4$: Deoxygenated blood flows from the brain to the heart.
$E_5$: Deoxygenated blood flows from the liver to the heart.
$E_6$: Deoxygenated blood flows from the gut to the heart.
$E_7$: Deoxygenated blood flows from the rest of body to the heart.
$E_8$: The heart process the oxygen and incoming deoxygenated bloods and generates *oxygenated* blood.
$E_9$: Oxygenated blood flows to the brain.
$E_{10}$: Oxygenated blood flows to the liver.
$E_{11}$: Oxygenated blood flows to the gut.
$E_{12}$: Oxygenated blood flows to the rest of the body.
$E_{13}$: The brain processes the oxygenated blood and generates deoxygenated blood.
$E_{14}$: The liver processes the oxygenated blood and generates deoxygenated blood.
$E_{15}$: The gut processes the oxygenated blood and generates deoxygenated blood.
$E_{16}$: The rest of body processes the oxygenated blood and generates deoxygenated blood.

The primary problem in this model is how to carve the static model into events. Note that we can make each generic action an event, but this would produce too many (unnatural) events.

Fig. 8 shows the chronology of events.

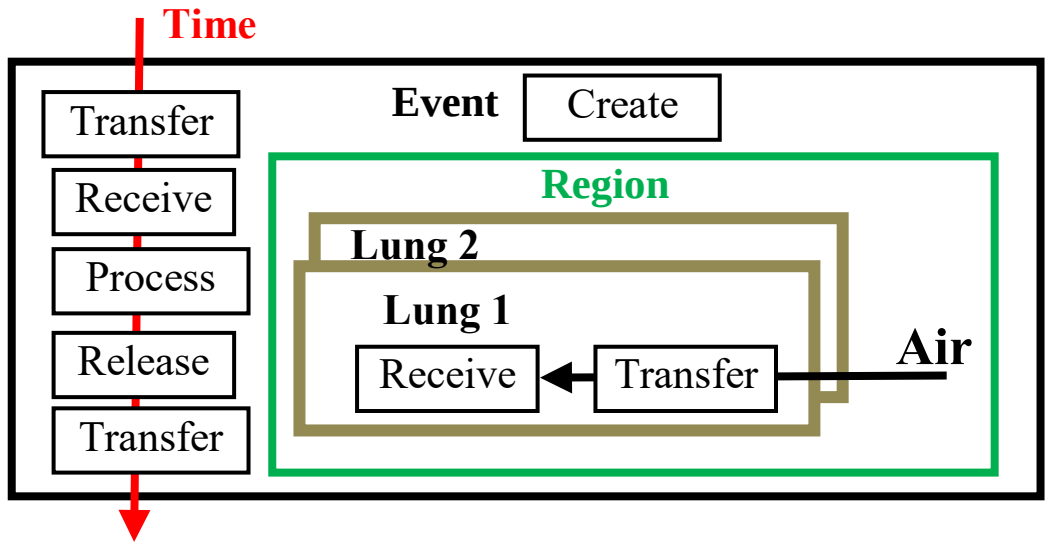

Fig. 6 The event *Air enters the lungs.*

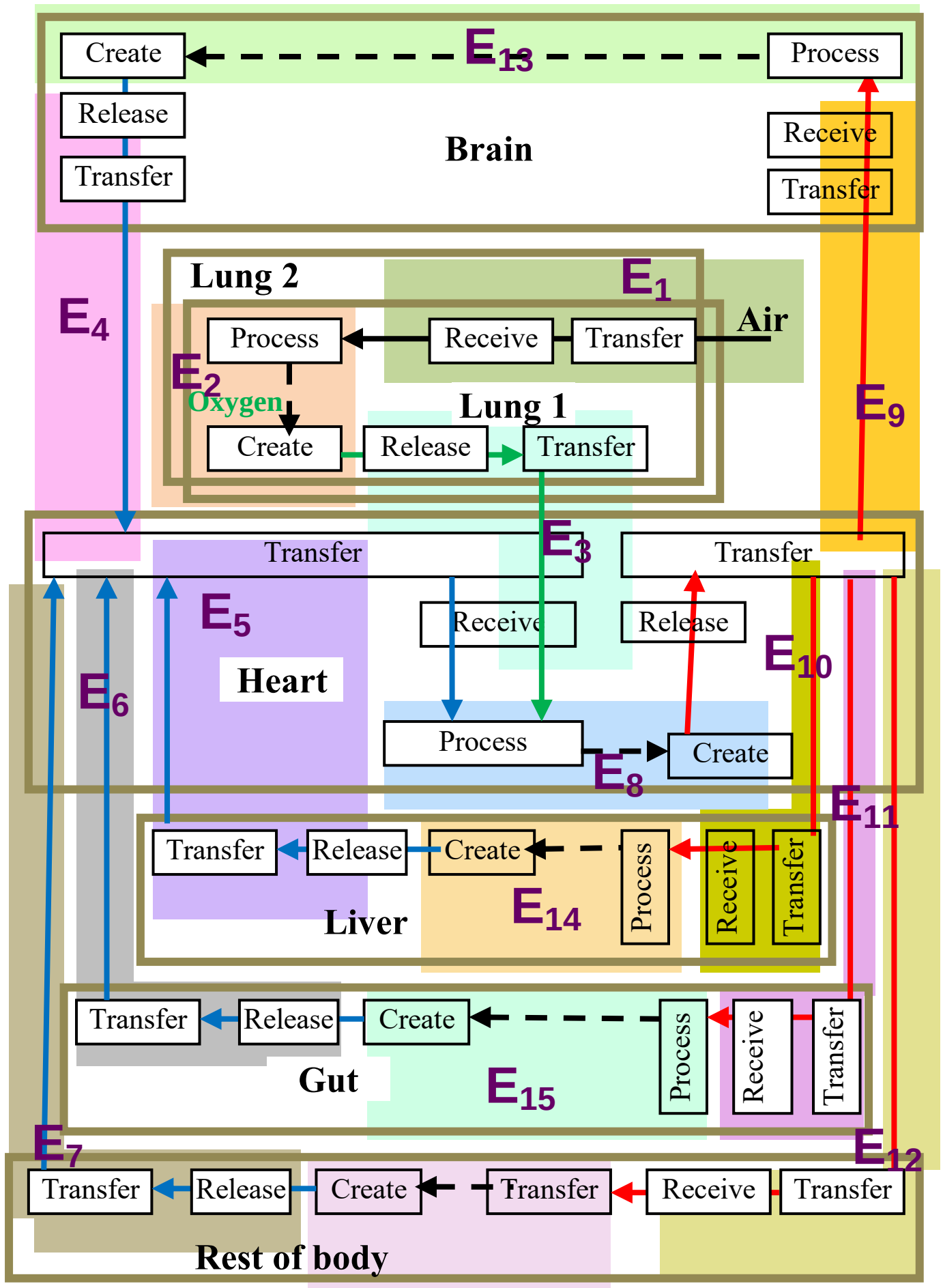

Fig. 7 Dynamic TM model.

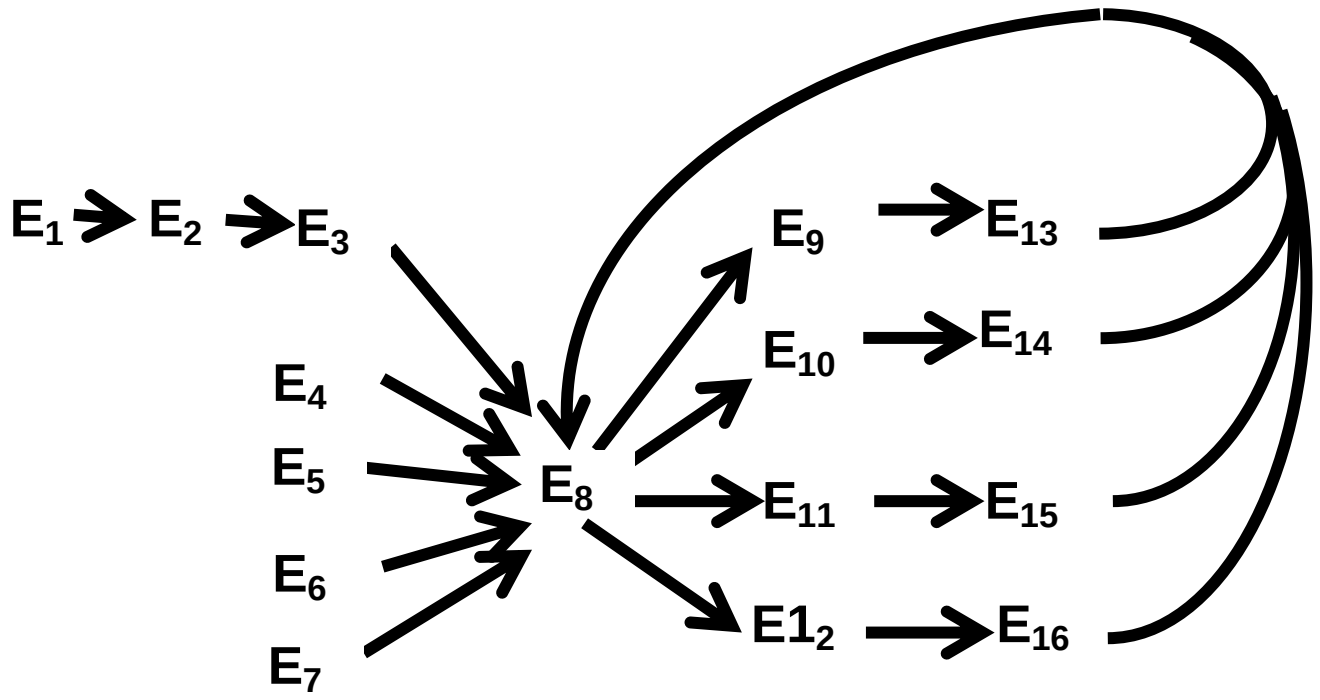

Fig. 8 Chronology of events (one inhalation).

*B. Slightly Confused Gork*

As described (and will partially corrected by the author) in **Gork**, the term "thinging machine" ("thinking" must be included in the request to Grok) refers to the conceptual framework in the context of software engineering and system modeling. It is a diagrammatic modeling language designed to represent systems through five fundamental actions: **create**, **process**, **release**, **transfer**, and **receive**. These actions describe how "things" flow and transform within a system, providing a unified way to model both static and dynamic aspects of systems. For example, in a coffee vending machine modeled as interconnected thinging machines of ground coffee, hot water and coffee thinging machines are as follows.

According to Gork, in a coffee vending machine modeled as a thinging machine,

- **Create**: A user inputs a *coin*, or payment.
- **Process**: The machine processes the payment and brews the *coffee*.
- **Release**: The machine dispenses the *coffee*.
- **Transfer**: The *coffee* is moved to the output tray.
- **Receive**: The user retrieves the *coffee*.

But Gork mixes "coin and coffee" TMs across "things" (user, vending machine). The following corrects this.

***Coins** TMs:*

- A user **creates**, **releases**, and **transfers** (output) coins to the vending machine.
- The vending machine **transfers** (input), **receives**, and **processes** the coins (to check that the coins make up the right amount); **releases** extra coins if the amount is more than the price; and **transfers** the extra coins to the user.

***Coffee-related** TMs:*

- The machine, which has ground coffee stored inside it, **releases and transfers** (output) the ground coffee to the brewer.
- The machine, which has hot water stored inside it, **releases and transfers** (output) the hot water to the brewer.
- The brewer **transfers** (input), **receives,** and **processes** both the ground coffee and hot water to **create**, **release**, and **transfer** (output) the coffee to the user.

Note that in a TM, a thing may be input but never received (e.g., inputting coins that are struck before they reach the coin-recognition mechanism) and a thing may not flow outside the machine even though is released (e.g., email is sent, but the communication channel is off).

This TM framework is particularly useful in fields such as software engineering, as it offers a clear, theory-based approach to design, unlike practice-based tools, such as UML. It aligns with the broader concept of *poiesis* (the study of creation and production) and can be applied to various domains, such as modeling propositional logic or workflows, as we've explored before.

## IV. TM Creation and existence: Ontological Dimension

Throughout the evolution of the TM model, each research paper has included a section that reflects an attempt to clarify and enhance the ontological foundation because the model involves metaphysical exploration of things in reality (see [12]).

Notions such as thimac, region, event, and reality are metaphysical concepts that need to be analyzed because such notions are intended to establish quality assurance though some objective foundation not subject to misapprehensions and conventions. Much metaphysical work might best be understood as a model-building process. Here, a model is viewed as a hypothetical structure that we describe and investigate to understand more complex, real-world systems. Accordingly, this section is an add-on to the previous discussion of this side of the TM model

*Creation* at the TM static level (subsistence) concerns a map of how thimacs appear in the world to be and possibly realized at the dynamic level (existence). To *exist* means to be a region that is processed in time (i.e., an event, to whatever times they exist). Only a thing that has a region can appear as an event (existing thimac), standing out from the multitudes of regions that compose the static background-as-stage of that appearance.

Everything that exists should be an event of some kind. The actualization process includes other regions (e.g., matter and energy to bring events into reality). The event is the "place" for processing existence (see the notion of existence container (*exicon*) [8]).

*A. Existence*

Generally, the notion of existence encompasses questions related to the concept of existence at large, where the TM model defines events as what it means to be an existing thing. An event is what "happens" in a region. Reality is viewed as a composite thimac with a singular unified totality.

Fig. 9 expresses the event *A thing exists*. Fig. 10 shows another representation of the same event in terms of region and time. Fig. 11 shows that the event by itself as a slice of existence of any region in time. A slice of existence can be abstracted from any region. This slice of existence will be called an existence container (exicon) collapsed from any particular event.

Fig. 12 shows an exicon as a slice of existence. Event(s) occupy exicons (individual existences); that is, they stand under or uphold events throughout existence as a phenomenal trait. The exicon is a TM non-thing agent that is responsible for the existence mode of events. It is a slice of existence by itself or what it is to be existent, analogous to the Aristotelian soul as what it is to be alive. There are numerous exicons coming into existence and going out of the realm of existence everywhere all the time. A deeper observation of the notion of existence is that regions, the very essence of thingness, are accidents of exicons. That is, regions are accidental to exicons as pure existence. Note the exicon is not accessible to human consciousness and is grasped only in an event's setting.

## B. *Exicon and Aristotle*

The TM event is a construct that weaves together (i) an actual existing specific region (a manifestation of beings) and (ii) the exicon, a slice of generic existence (a manifestation of being): "this" (a slot of being) (see Heidegger's conception of metaphysics [11]). An event obviously includes (a) a *being* manifestation of a thimac (and its region is part of this being) and (b) an exicon.

Figs. 13 and 14 show these constitutions of the event. The constitutions may be just a blend of the two elements. The exicon (piece of Being) is definitely not a thimac because it lacks the *create* action; it is pure existence that does not need creation.

The issue here may be related to the classical notion of categories. Categories such as quality, relation, posture, and place presuppose something that it predicts. Aristotle calls this subject to which all categories apply *substance* [13]. Substances are

(1) The simple *bodies* (i.e., earth, fire, and water and the things composed of them). Substances are those things that best merit the title "*beings"* [14]. This is usually interpreted to mean those things that are the foundational or the basic things from which everything is constructed. Some thinkers interpreted impressions and Plato's forms as substances for the same reason.

(2) That which is present in such things is the cause of their being as the *soul* is the being of animals.

(3) The parts that are present in such things.

(4) The essence [13].

It follows, then, that the substance has two senses,

(a) The ultimate substratum, which is no longer predicated of anything else, and

(b) That which is a "this" and separable—and of this nature—is the shape or form of each thing [13].

If we project these notions in the TM exicon, we can say that what we called the *manifestation of Being* (Fig. 13) that which includes (2) and *the manifestation of being* (Fig. 14). This may relate to Heidegger's distinction between being "opposed to" or "against" *becoming*, that is, the notion "to be" present (in the sense of *being* here or there and "to become," or "to come into the specific forms of beings, or to emerge, to appear as beings and maintain their presence (Being))" [11].

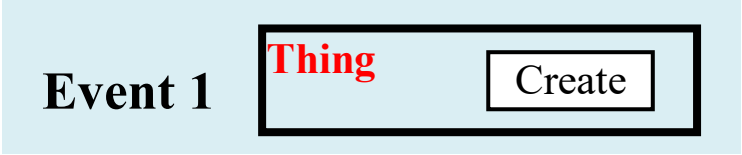

Fig. 9 *A thing exists.*

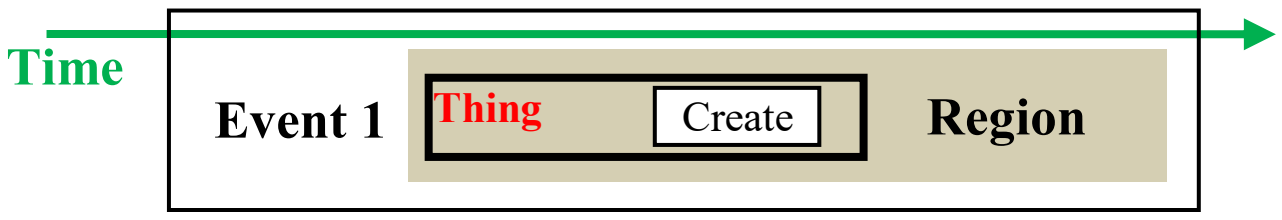

Fig. 10 Another representation of an event.

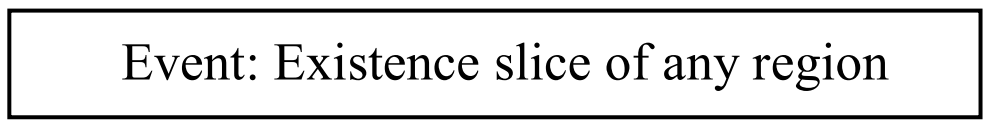

Fig. 11 Existence slice of any region.

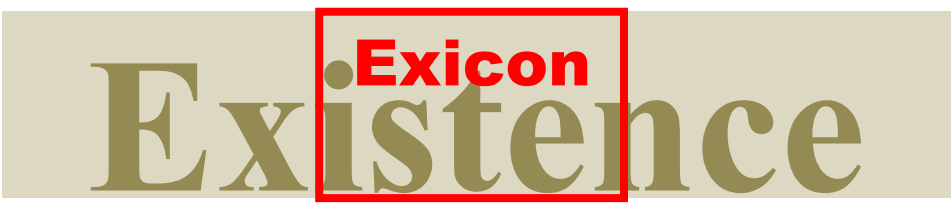

Fig. 12 Capsule of existence.

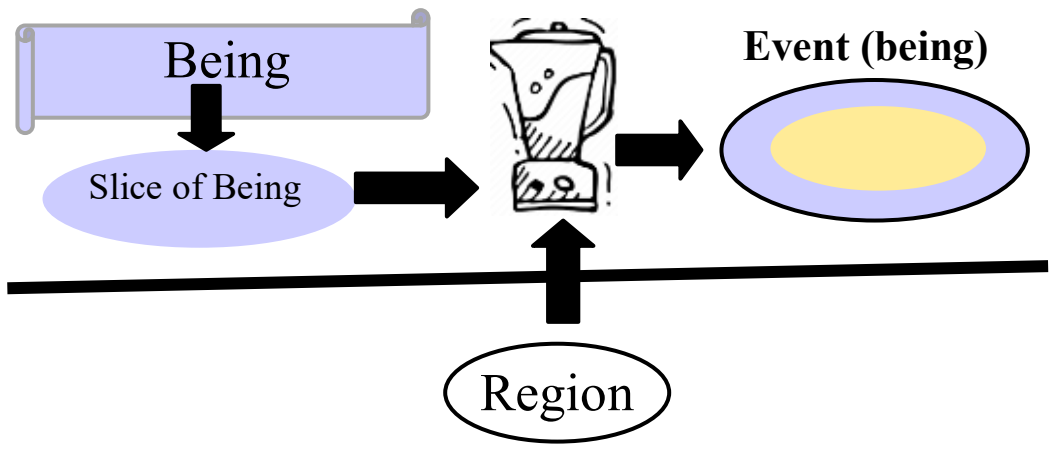

Fig. 13 Illustration of TM view of existence.

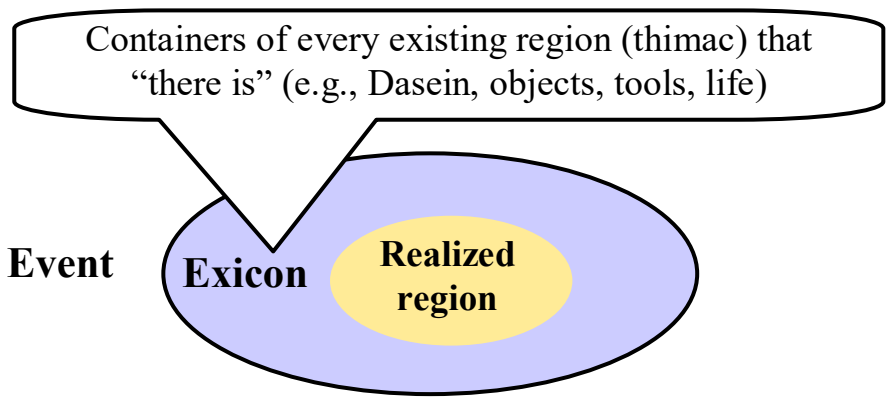

Fig. 14 Illustration of TM view of existence.

According to [11], Heidegger observed *being( )becoming* with dynamics of the two halves making a whole, as shown in Fig. 14 (top).

Such a discussion points to future research that links the TM to philosophical concepts.

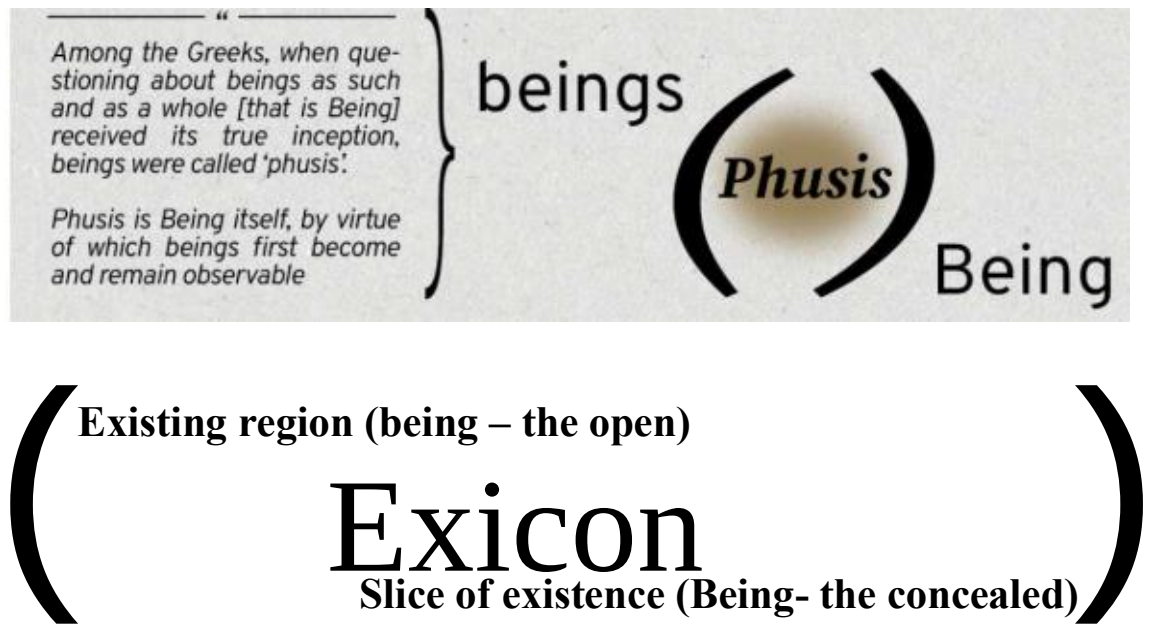

Fig. 15 *Being( )becoming* as illustrated by [11] (top) and the corresponding TM structure (bottom).

## V. CARVING NATURE

What kind of conceptual differences count for showing significant differences in conceptual systems? Note that a conceptual model is a formulation of conceptual system. According to [2], “Philosophers have assumed that there is only one kind of difference: how a conceptual system “carves up nature.” In one view, a conceptual system can succeed or fail to fit well, that is, to “carve nature at the joints.” According to Lewis (see [15]), there is an objective difference between natural and unnatural properties that carve reality at the joints, and what makes properties (e.g., mass, charge, quark color) natural has to be something about the intrinsic nature of reality [15]. We can choose various joints to carve at; that is, we can conceptualize various aspects of reality. In addition, conceptual systems can vary in their “fineness of grain”; that is, they can carve nature into big chunks or small artful ones [2].

However, a conceptual system cannot create new joints because it is assumed that all the joints are given ahead of time. A conceptual system is accurate if it always finds joints in nature (though it certainly won’t find all of them) and inaccurate if it misses the joints and hits a bone or nothing at all [2]. However, such a view leaves out concepts that are not to be found objectively in nature but which are a result of the human imaginative capacity [2].

A Taoist allegory offers some advice. The king asks about his butcher’s impressive knife work. ’Ordinary butchers,’ he replies, ‘hack their way through the animal. Thus their knife always needs sharpening. My father taught me the Taoist way. I merely lay the knife by the natural openings and let it find its own way through. Thus it never needs sharpening’” ([4], referencing [16]).

In TM, carving at the joints bottoms out on the base level of the generic actions, create, process, release, transfer, and receive. For example, the flow of a thing from one thimac to another supervenes on lower-level structures, such as send and receive, which in turn supervene on the lower-level atomic structure of the genetic actions release, transfer, and receive (see Fig. 1 in the introduction). The generic actions determine a single way that the world is.

*Example*: [15] provided an illustration of a portion of reality, a 9x9 grid (Fig. 16), and asked what the boundaries that matter intrinsically are. The obvious way to carve is the one demarcating the 81 squares, or the grid can be divided horizontally (i.e., rows), vertically (i.e., columns) or in other ways, as shown in Fig. 16, using the TM create action.

Suppose a few things (letters) appear in the previously blank grid. The end result is depicted in Fig. 17 (left). These things may be interpreted as dividing the grid into “square pairs,” each comprising two adjacent squares containing an X/O pair (Fig. 17 – right). The single X in the upper left part of the grid may be interpreted as an anomaly [15].

Suppose a grid includes letters, shown in Fig. 18 (left). This arrangement of X’s and O’s is interpreted as tic-tac-toe as shown in the figure. It turns out that only the borders around the 9 squares in bold in the converted grids above really mattered, as shown in Fig. 18 (right)[15].

In a TM, the dividing operation is not limited to static structures. The carving problem also involves the dynamic structures. For example, in Fig. 19, carving may be associated with *processes* that move things from one thimac to another, processes the changes things, or processes that move things outside the modeled domain. In general, there are two types of carving, structural and processual, as will be shown in the next section. We claim that the recognition of these types of nonentities based carving is unique to the TM model.

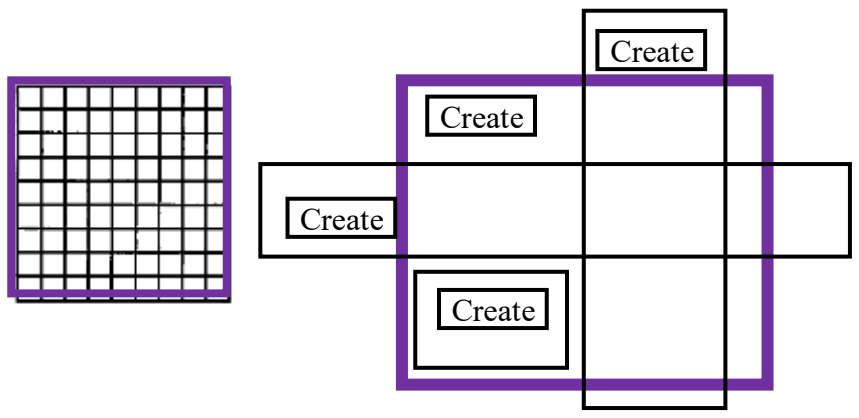

Fig. 16 Grid (left) and samples of divisions (right).

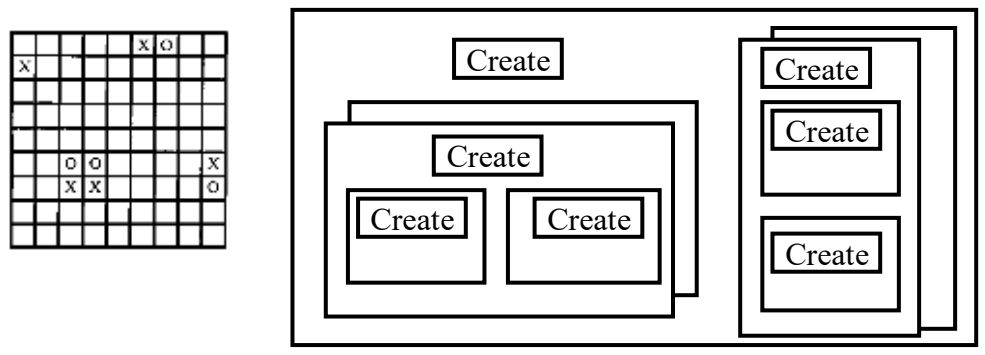

Fig. 17 The grid with things (left) and their TM representation as two adjacent squares (right).

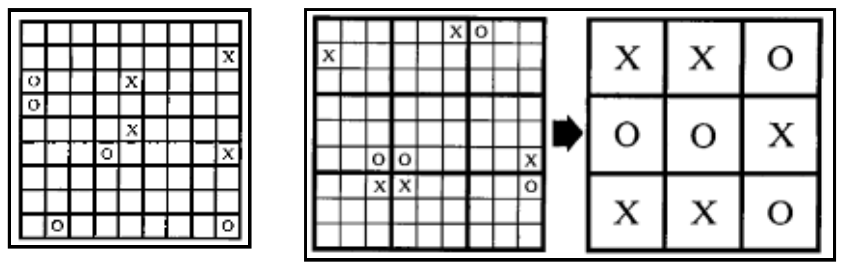

Fig. 18 Another carving of the 9x9 grid.

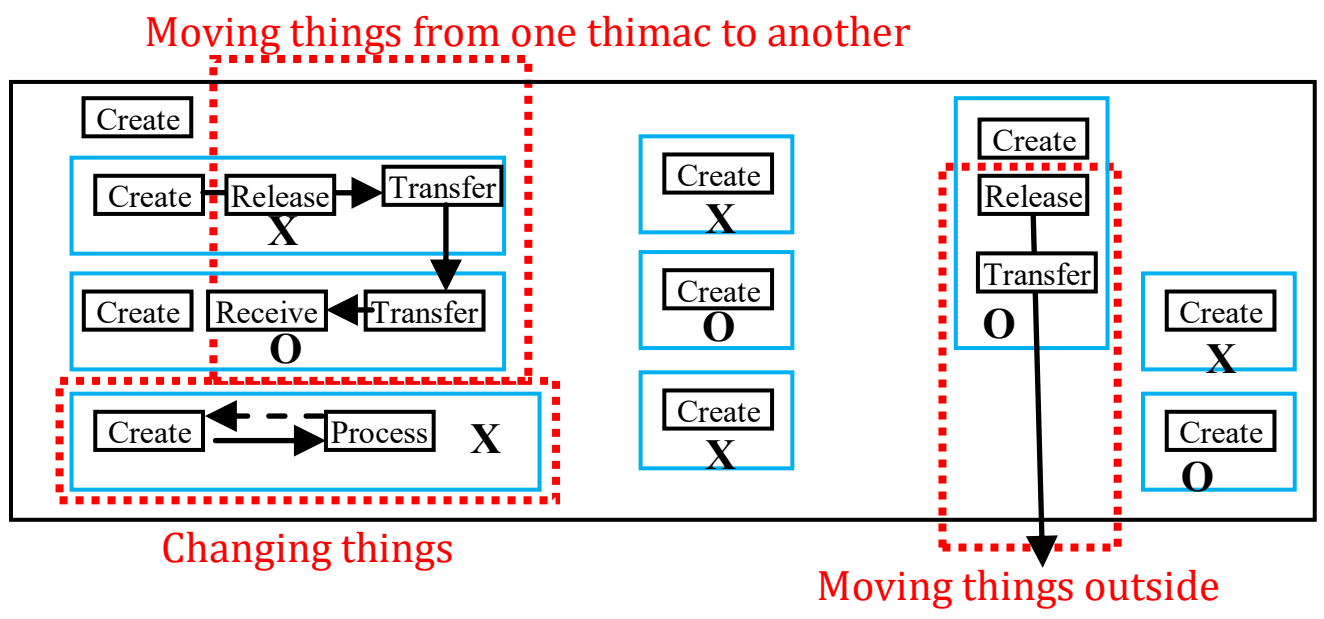

Fig. 19 The TM model introduces dynamic carving.

## VI. Contrasting with Typical Modeling

The entity/relationship (ER) model is widely used as a conceptual formalism to provide standard documentation for relational information systems. The aim of this section is to contrast the ER modeling techniques with TM modeling, especially with respect to carving at the structural and processual levels. The example demonstrates how the TM model lends itself to the carving process by separating the static/dynamic models. In the next example, for simplicity's sake, some *create* actions are removed under the assumption that the box is sufficient to indicate the presence of the thimac.

[17] presented nine requirements, each of which is exemplified by a real-world clinical case study. The data model supports explicit hierarchies, multiple hierarchies, and non-strict hierarchies in dimensions. The model handles change over time and is equipped with algebra. The involved case study concerns the patients in a hospital, their associated diagnoses, and their places of residence, as shown partially in Fig. 20.

Fig. 21 shows the structural TM model carving that includes the thimacs patient (number 1 in the figure), area (2), and diagnoses (3) and *Has* (4) as the relationship between apatient member (5) and a diagnosis subset (6).

Without loss of generality, to simplify the diagram in Fig. 21, we limit the number of attributes in the given example (Fig. 20), and we assume that SSN and code are the keys of patient and diagnosis, respectively. The relation *Has* is specified to include one member of patient (5) and a subset of diagnoses (6). Note that *Has* is a thimac and patient and diagnosis are subthimacs (participants) in *Has*. The *Has* thimac can be specified as flows (release, transfer, transfer, receive) of diagnosis to *patient*.

The structure of thimacs in Fig. 21 is a design decision. For example, it is possible to make *area* a sub-thimac (attribute) of patient. Furthermore, Fig. 21 can be simplified by removing *creates* due to the assumption that the boxes indicate the appearances of thimacs in the model, as shown in Fig. 22.

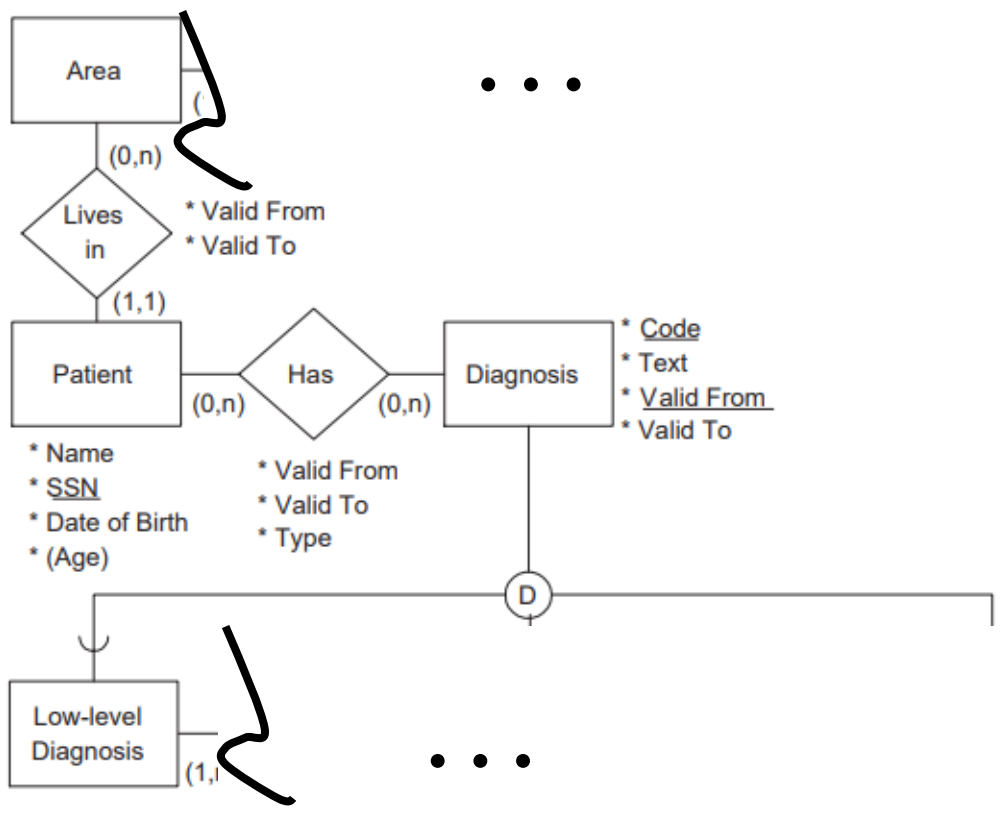

Fig. 20 Case study model (partial picture, from [17])

### A. *Static TM Model*

The complete static model includes the previously described structural framework and all of the thimacs' actions. We will just describe the procedure required to *insert a new patient in the database and insert a new diagnosis of a patient in the database*. Fig. 23 shows such a model.

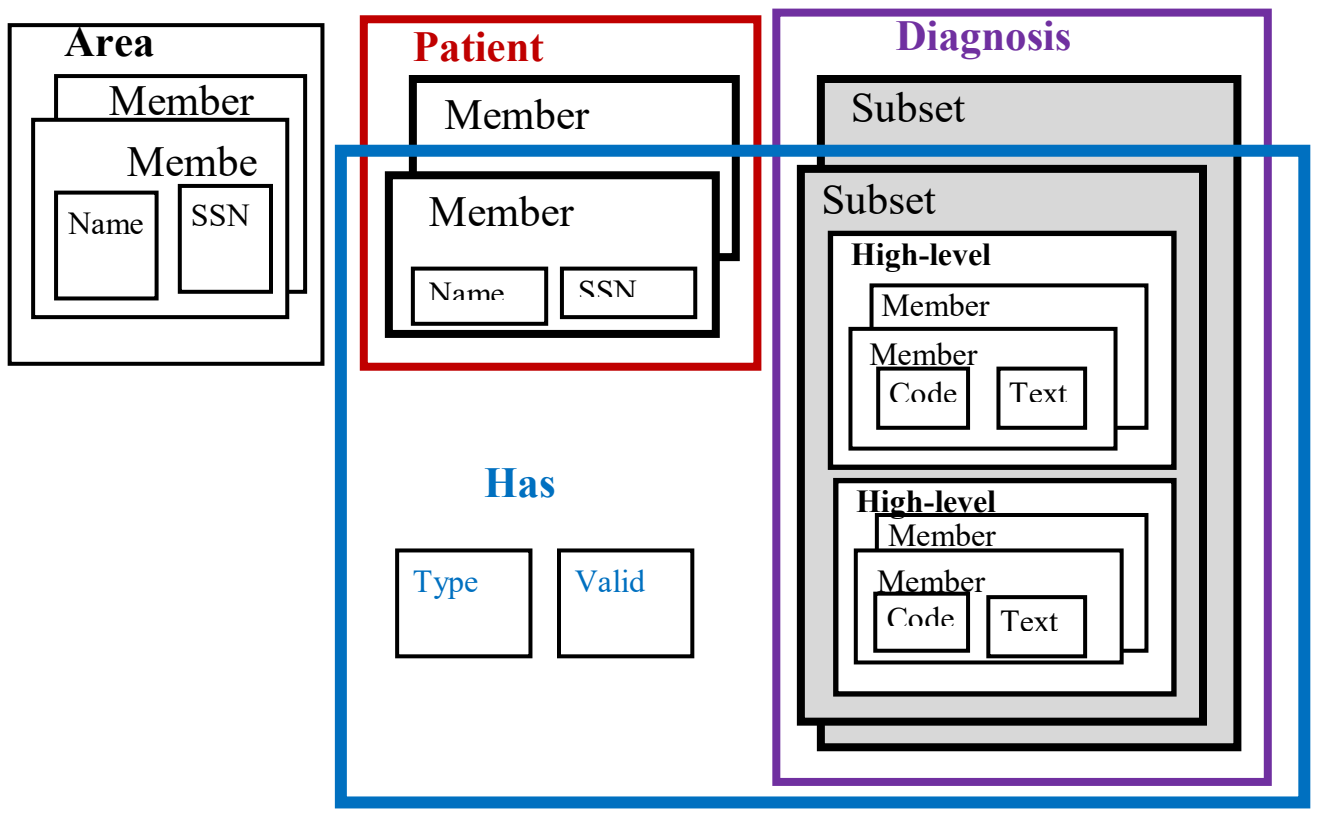

Fig. 22 Simplified structure of the TM model.

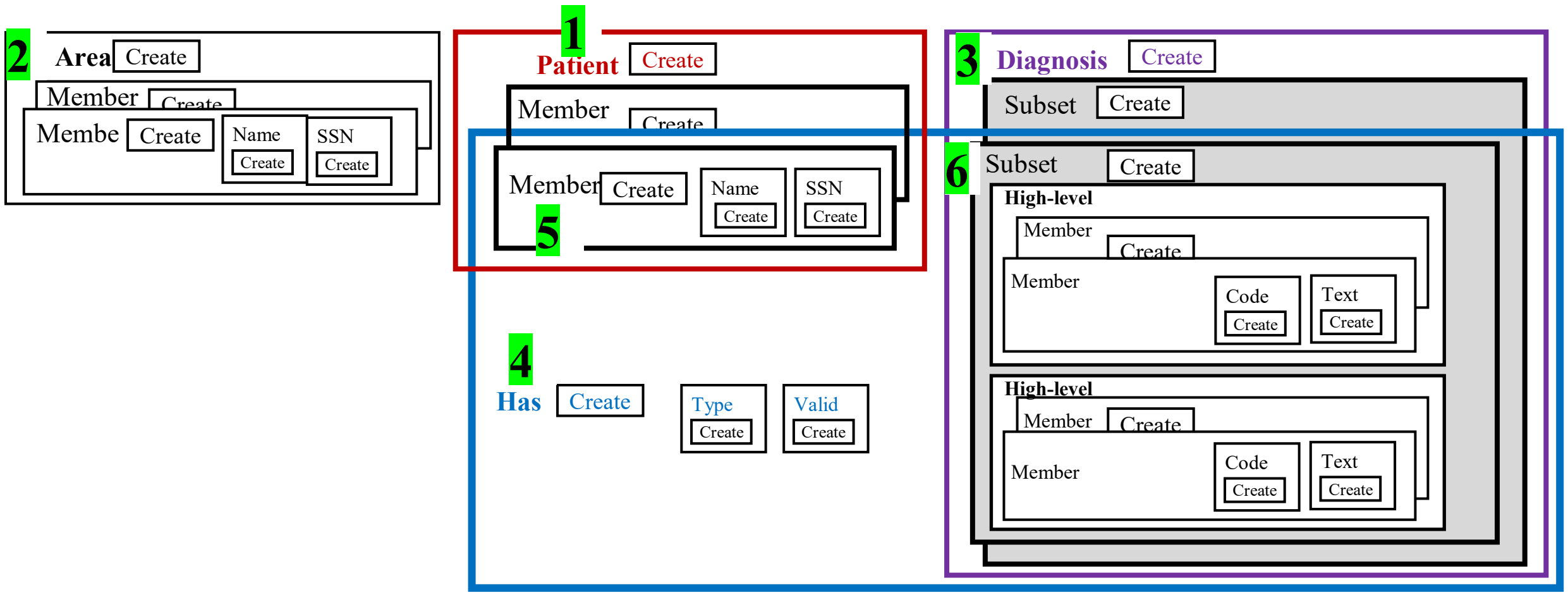

Fig. 21 The structure of the TM model.

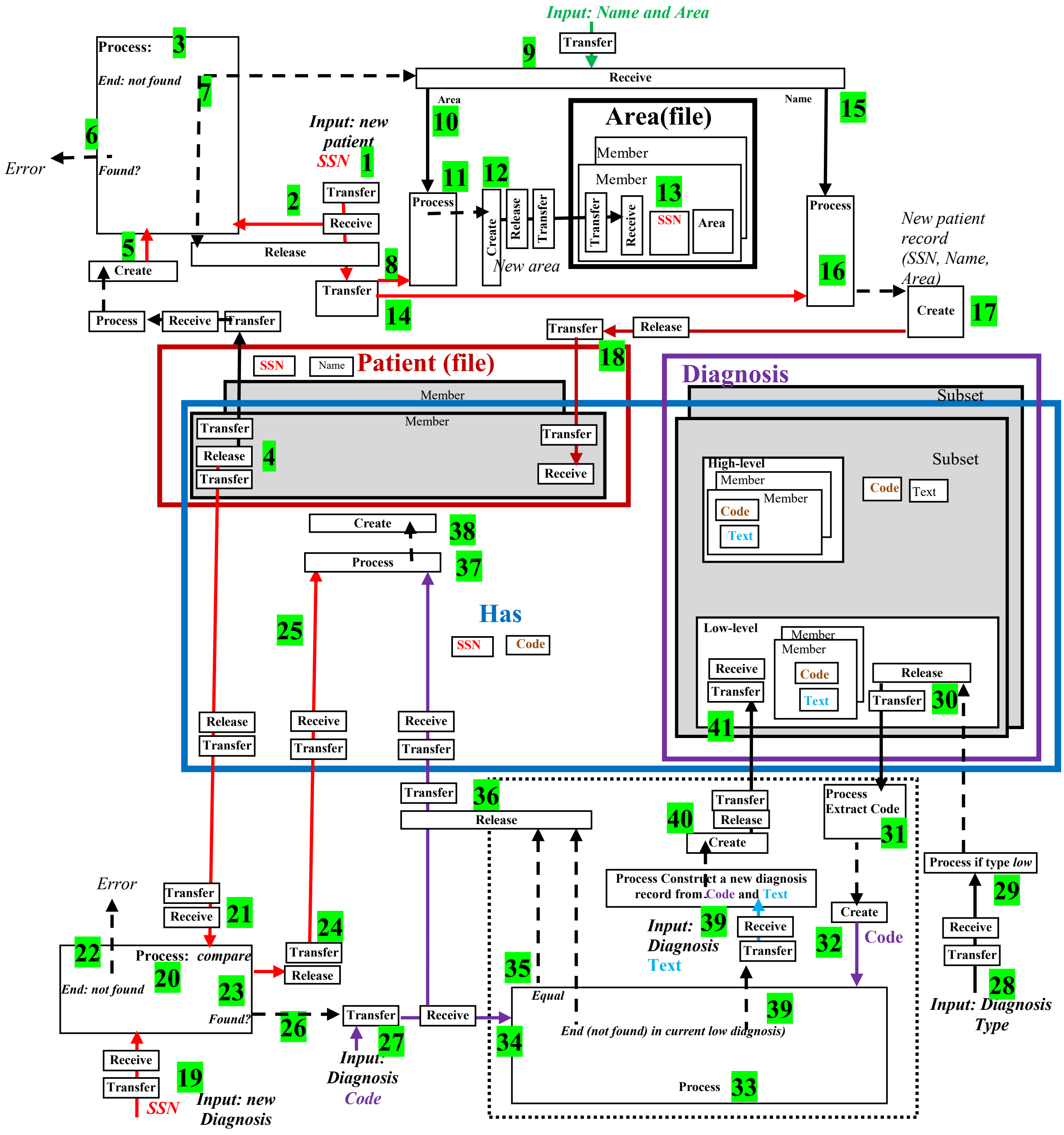

Fig. 23 TM static model inserting a new patient in the database and inserting a new diagnosis of a patient into the database.

**Inserting a new patient**

1. The new patient's SSN is inputted.
2. The SSN is processed (2 and 3) to be compared with the current SSN in the database (4 and 5). Note that to get this SSN, a patient record is retrieved (4) and processed to extract its SSN. The *create* (5) here indicates the appearance of the SSN at the global level after it was embedded in the patient record.
3. The comparison between the inputted and extracted SSNs is realized through a loop through all patient records that will be specified in the dynamic model. Without loss of generality, we assume here a sequential search.
4. The new patient's SSN is inputted.

5. The SSN is processed (2 and 3) to be compared with the current SSN in the database (4 and 5). Note that to get this SSN, a patient record is retrieved (4) and processed to extract its SSN. The *create* (5) here indicates the appearance of the SSN at the global level after it was embedded in the patient record.
6. The comparison of the input and extracted SSNs is realized through a loop through all patient records that will be specified in the dynamic model. Without loss of generality, we assume here a sequential search.
7. If the SSN is found in the database, then an error message is printed (6).
8. If the SSN is not found (7), then it is released (8) to construct (a) a new record of area along with inputted *area* value and (b) a new record of patient along with inputted *name*.
9. To construct a new area record, the SSN (8) and inputted *area* value (9 and 10) are processed (11) to create a record (12) that is inserted in the area file (13).
10. To construct a new patient record, the SSN (14) and the additional data, the name (15), are processed (16) to create a new patient record (17), which is inserted into the patient file (18). As we mentioned before, without the loss of generality, the patient attributes are limited here to *name*. Note that here, we are simplifying the process. It is possible here to input the *name* and the *area* separately.

**Inserting a new patient diagnosis**

1. The patient's SSN is inputted (19).
2. The SSN is processed to be compared with SSNs currently in the database to establish that the given SSN is in the current database (20 and 21).
3. If the inputted SSN is not in the database, an error is given (22).
4. If the inputted SSN is found (23) in the database, the SSN is sent to *has* to construct, later, the relationship record (24 and 25). Additionally, the data of the diagnosis *code* is triggered (26) to be inputted (27) to complete, later, the relationship *has* record.
5. Accordingly, the *type* of diagnosis is triggered to be inputted (28). We assume that the type is "low" (29). Now, the dotted box represents the process of establishing whether the diagnosis is already in the diagnoses file and constructing the relationship *has*, updating the diagnosis file accordingly.
6. Because the *type* of diagnosis is low diagnosis (28), we start a loop to check whether the diagnosis is already in the file by retrieving a record from the diagnosis file (30). The record is processed (31) to extract its code (32).
7. The extracted code is processed (33) by comparing it with the inputted code (34). If the two codes are equal (35), then this means that it is not necessary to create a new diagnosis record. Accordingly, the code is sent to the relationship *has* (36) to construct a record of the relationship *has* (37 and 38).
8. If the end of diagnosis file is reached, then in addition to establishing the relationship *has*, an input is triggered (39) to provide the new diagnosis (39), which is processed along with the code to create a new record in the diagnosis file (40 and 41).

B. *Dynamic Model*

Accordingly, we overlay the dynamic model on the static model, in which events are specified in a sub-diagram. An event is a thimac in time. For example, Fig. 24 shows the event *Inputting a new patient to the database*. The region in the figure is a sub-diagram of the static model. However, for simplicity's sake, we will represent the event by its region.

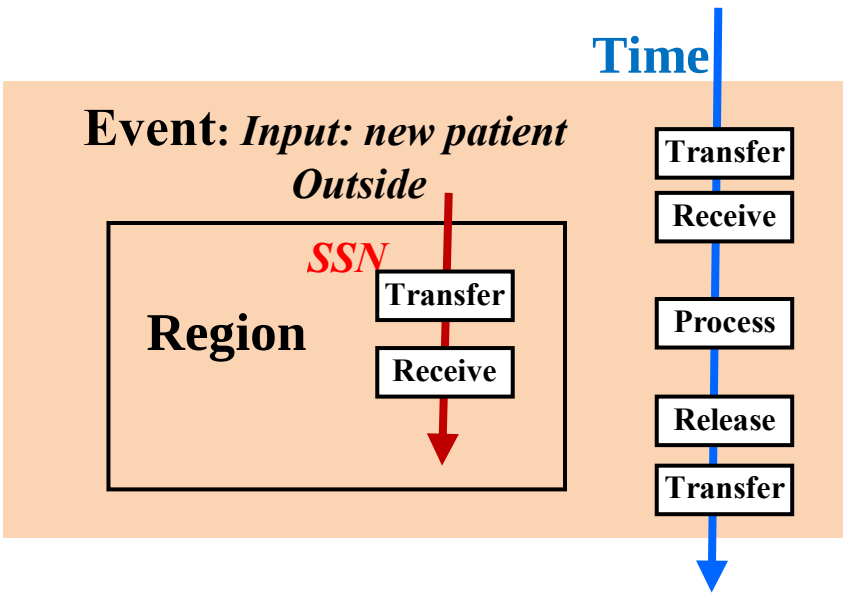

Fig. 24 The event *Inputting a new patient to the database.*

Accordingly, the set of events in the dynamic model is specified in Fig. 25 as follows.

**Inserting a new patient**

$E_1$: Inputting an SSN.
$E_2$: Retrieving a record from the patient database and extracting its SSN.
$E_3$: Comparing the inputted and retrieved SSNs.
$E_4$: The SSN is found in the patient file, and an error occurs.
$E_5$: End of patient's file; therefore, data about the new patient is requested and inputted.
$E_6$: A record (SSN, area) is constructed and inserted into the area file.
$E_7$: Record (SSN, name) is constructed and inserted in the patient file.

**Inserting a new diagnosis of a patient**

$E_8$: Inputting an SSN.
$E_9$: Retrieving a record from the patient database and extracting its SSN.
$E_{10}$: Comparing the inputted and extracted SSNs.
$E_{11}$: End of patient file. Error.
$E_{12}$: SSN is found in patient file.
$E_{13}$: Code of the diagnosis is requested and inputted.
$E_{14}$: The type of diagnosis is requested and inputted. We assume it is of the low type.
$E_{15}$: A low record is extracted from the low file, and its code is extracted.
$E_{16}$: The inputted and extracted codes are compared.
$E_{17}$: The inputted and extracted codes are equal (the diagnosis is already in the diagnosis file).

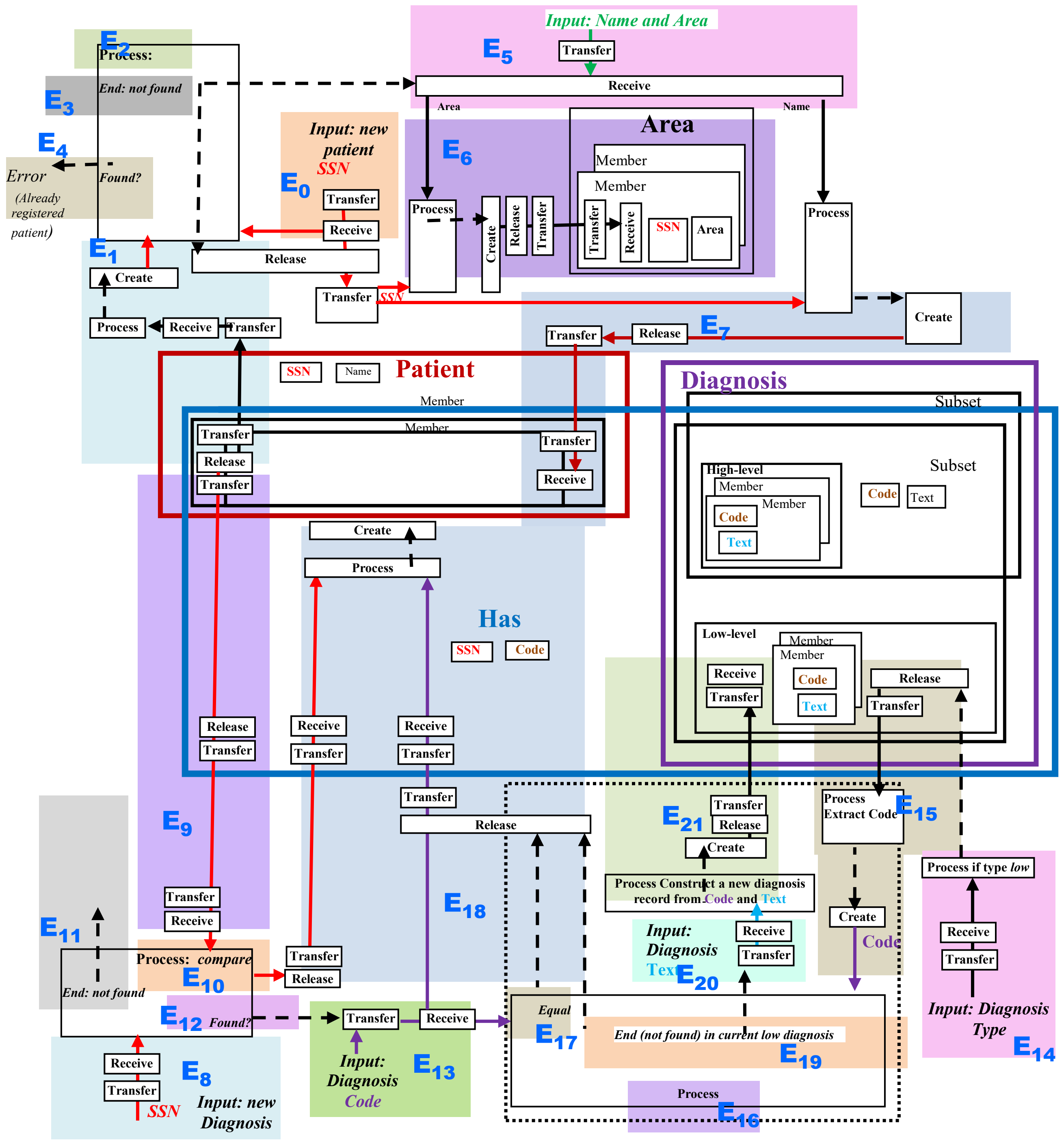

Fig. 25 The Dynamic model

$E_{18}$: A new record (SSN, code) is constructed and inserted in the relationship *has*.

$E_{19}$: The end of the diagnosis file. The input code does not correspond to any stored diagnosis.

$E_{20}$: Requesting and inputting a new diagnosis.

$E_{21}$: Construct the record (code, text) and insert it into the low diagnosis file.

Fig. 26 shows the chronology of events in the system.

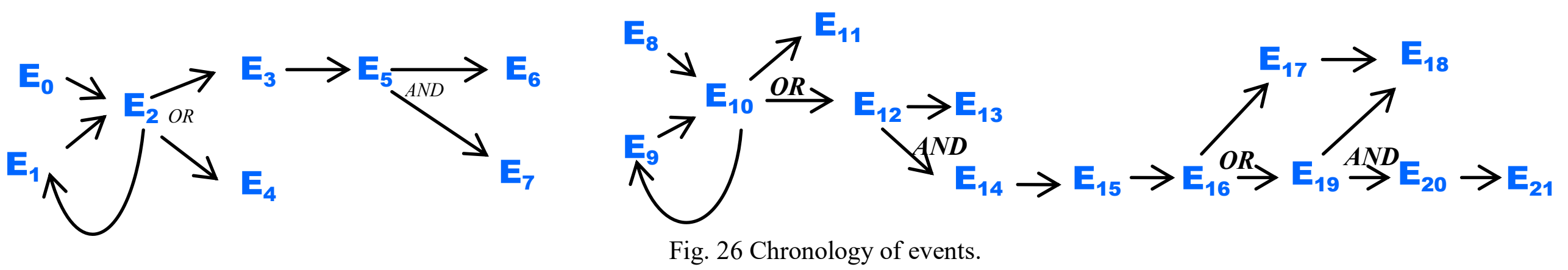

Fig. 26 Chronology of events.

## Conclusion

The carving metaphor has been used to build a conceptual system of reality. This paper focuses on applying the notion of carving to the TM model. The central problem is how to carve events when building a TM model. A TM event is defined as a thimac with a time breath that infuses dynamism into a static region. A region is a diagrammatic description based on five generic actions: create, process, release, transfer, and receive. The carving problem includes structural carving and dynamic events. The focus is on the events in which the classical treatment is centered on categorization.

The diagrammatic language of the TM model facilitates identifying and classifying of processes based on the five actions. The paper contains new material about TM modeling and generalizations and focuses on the carving problem to include structural carving and dynamic events. The study's results provide a foundation for establishing a new type of carving reality based on the TM model diagrams. The study opens up a new direction in the design of conceptual model using a thimac.

## References

[1] N. Guarino and G. Guizzardi, "We need to discuss the relationship: revisiting relationships as modeling constructs," *Lect. Notes Comput. Sci.*, vol. 9097, 279-294, 2015. https://doi.org/10.1007/978-3-319-19069-3_18.

[2] G. Lakoff, *Women, Fire, and Dangerous Things: What Categories Reveal about the Mind*, Chicago: The University of Chicago Press, 1987. https://lecturayescrituraunrn.wordpress.com/wp-content/uploads/2017/03/unidad-5-lakoff-women-fire-and-danger.pdf

[3] A. Boldachev, "Event ontology vs object ontology," Medium blog, June 20, 2023. https://medium.com/@boldachev/event-ontology-vs-object-ontology-cef764feb12c

[4] M. H. Slater and A. Borghini, "Introduction : lessons from the scientific butchery," in Carving Nature at its Joints: Natural Kinds in Metaphysics and Science, J. K. Campbell, M. O'Rourke, and M. H. Slater, Eds. Cambridge: MIT Press, 2011.

[5] L. A. Paul, A one category ontology, in Being, Freedom, and Method: Themes from the Philosophy of Peter van Inwagen, J. A. Keller, Ed, Oxford Scholarship Online, January 2017, https://lapaul.org/papers/Paul-OneCategory.pdf

[6] C. H Calisher, "Taxonomy: what's in a name? Doesn't a rose by any other name smell as sweet?" *Croat Med J.*, vol. 48, no 2, 268-270, April 2007.

[7] C. Strößner, "Criteria for naturalness in conceptual spaces," *Synthese*, vol. 200, no, 78, 2022.

[8] S. Al-Fedaghi, "Preconceptual modeling in software engineering: metaphysics of diagrammatic representations," HAL archives, 2024. https://hal.archives-ouvertes.fr/hal-04576184.

[9] T. Morton. *The Ecological Thought*. Cambridge: Harvard University Press, 2010.

[10] "Biology Notes for IGCSE," blogspot.com, 2014. https://biology-igcse.weebly.com/the-circulatory-system.html

[11] A. C. Rollino, "Being as place: introduction to metaphysics – part two (the limitation of being)," April 16, 2022. https://rethinkingspaceandplace.com/2022/04/16/being-as-place-introduction-to-metaphysics-part-two-the-limitation-of-being/

[12] S. Al-Fedaghi, "Exploring conceptual modeling metaphysics: existence containers, Leibniz's monads and Avicenna's essence," arXiv preprint, arXiv:2405.01549, February, 20, 2024.

[13] A. P. Tabifor, V. B. Ngalim, and N. Shang, "Substance: the foundation to Aristotle's metaphysics," *Social Science and Humanities Journal*, vol. 8, no. 3, March 2024.

[14] H. Robinson and R. Weir, "Substance," in The Stanford Encyclopedia of Philosophy (Summer 2024 Edition), E. N. Zalta and U. Nodelman, Eds. https://plato.stanford.edu/archives/sum2024/entries/substance/.

[15] S. Page, "Carving nature at its inherent joints: the problem of the independent criterion," *Auslegung: A Journal of Philosophy*, vol. 29, no. 1, fall/winter 2007.

[16] M. Kahn, *The Tao of Conversation*, Oakland: New Harbinger, 1995

[17] T. B. Pedersen and C. S. Jensen, "Multidimensional data modeling for complex data," Proceedings 15th International Conference on Data Engineering, Sydney, NSW, Australia, March 23-26, 336-345, 1999.